# Tantalum nitride nanotube photoanodes: establishing a beneficial back-contact by lift-off and transfer to titanium nitride layer

Lei Wang,[a] Anca Mazare,[a] Imgon Hwang,[a] and Patrik Schmuki[a,b,*]

a Department of Materials Science and Engineering, WW4-LKO, University of Erlangen-Nuremburg, Martensstrasse 7, D-91058 Erlangen, Germany.
b Department of Chemistry, King Abdulaziz University, Jeddah, Saudi Arabia
* E-mail: schmuki@ww.uni-erlangen.de

Abstract

In this work we introduce the use of $TiN/Ti_2N$ layers as a back contact for lifted-off membranes of anodic $Ta_3N_5$ nanotube layers. In photoelectrochemical $H_2$ generation experiments under simulated AM 1.5G light, shift of the onset potential for anodic photocurrents to lower potentials is observed, as well as a higher magnitude of the photocurrents compared to conventional $Ta_3N_5$ nanotubes (~0.5 V RHE ). We ascribe this beneficial effect to the improved conductive properties of the $TiN_x$-based back contact layer that enables a facilitated electron-transport for tantalum-nitride based materials to the conductive substrate.

Key words: tantalum nitride; nanotube membranes; titanium nitride; water splitting

# 1. Introduction

Ta3N5 is considered to be one of the most important semiconductor materials for photoelectrochemical (PEC) water splitting. It has a visible-light active band gap of ~2.1 eV that embraces the H2O H2/O2 reduction-oxidation potentials, and thus may yield a 15.9% theoretical maximum solar-to-hydrogen (STH) conversion efficiency [1-9]. Extensive research has focused on nanostructuring the Ta3N5 photoelectrodes, in the form of nanoparticles, nanorods, or nanotubes [10-15]. 1-dimensional (1D) nanorod/nanotube structures provide the advantage, not only of a high surface area, but also of a directional and controllable light and electron management compared to the nanoparticles. For the synthesis of Ta3N5 nanostructures, first usually Ta2O5 structures are synthesized, that are then converted to Ta3N5 by a thermal annealing treatment in NH3. However, thick nanoparticle-based Ta3N5 photoelectrodes (e.g. deposited on conductive layers (FTO) by electrophoretic deposition) suffer from an inefficient electron transfer across the interface between the Ta3N5 and the conductive substrate [10,14]. To overcome this issue, a refined necking treatment (i.e. additional TaCl5 decoration and annealing) was introduced and improves the charge transfer not only between particles but also at the particle/substrate interface [10,11]. Recently, excellent mechanical and electric contacts were reported by using a transferred particle layer on a Ti/Ta-metal contact [12,13]. More recently a film transfer method based on Ta2O5 films on silicon for the Ta3N5 photoelectrodes was reported to achieve control in film thickness and to establish a defined back contact [14]. These data demonstrate that various metallic (Nb, Ti, Ta) and semiconductor (NbNx, CdS) layers can be used as back contacts to improve the PEC performance of Ta3N5 photoelectrodes.

For Ta3N5 photoelectrodes grown on or from Ta foils, either 1D nanostructures such as nanorods and nanotubes or compact films, even a higher PEC water splitting performance was reported compared with nanoparticle-based photoelectrodes. Also for these structures back contacts are a crucial factor for the quality of electrode-substrate connection [7,15-20]. Previous work has shown that the formation of Ta-subnitride layers (Ta2N, TaN) underneath anodic Ta3N5 nanotubes can enhance their efficiency due to a higher conductivity than Ta3N5, and therefore improve the charge transfer to the back contact [7,15]. However, to build photoanodes based on nanotubular Ta3N5 and explore different back contact layers, formation of free-standing membranes is needed. Similarly to the procedure for obtaining TiO2 nanotube membranes, under optimized conditions Ta2O5 nanotube membranes can be obtained, that can then be converted to Ta3N5. The methods used for obtaining membranes include either voltage pulses or an annealing-anodization-etching sequence [21] to separate the layer from the Ta metal substrate. In the present work, we use the latter, i.e. after nanotube growth, the layers are crystallized (air annealing), anodized again to grow an amorphous nanotubular oxide layer underneath, which then can be selectively dissolved in a HF aqueous solution. This procedure leads to Ta3N5 nanotube membranes and thus enables the possibility of transferring them to desired conductive layers as back contacts for 1D Ta3N5 nanotubular structures. Ideal ohmic back contacts should be highly conductive and have a work function corresponding to the Fermi level of the semiconductor at flatband situation.

For this we use in the current work TiN/Ti2N layers as a back contact. They seem ideal as they show a virtually metallic behavior and provide an excellent match of the work functions (W F (TiN x ) ≈ 3.5-4.4 eV) [22]. Here we show that indeed, by using this back contact, the anodic photoresponse under simulated AM 1.5G light shifts to significantly lower onset potentials. We ascribe this to the improved charge transfer properties of the back contact layer that enables a facilitated electron-transport to the substrate.

## 2. Experimental

The Ta2O5 NTs were prepared by anodizing Ta foil (99.9%, 0.1 mm, Advent) in a two electrode electrochemical cell with a Pt counter electrode. The anodization experiments were performed at 60 V (with a maximum current density set at 0.1 mA cm $^{-2}$ ) in a sulfuric acid (H2SO4 , 98%) electrolyte containing 0.8 wt.% NH 4 F and 13.6 vol% DI water. The anodization times were 5, 10, and 15 min, respectively. Then, samples were immersed in ethanol for 5 min and dried in N2 . For producing Ta2O5 nanotube membranes, the as-prepared Ta2O5 nanotube layers were annealed in air at 450 °C for 1 h, followed by anodizing at 80 V in the same fresh electrolyte. The layers were then detached from the Ta substrate by immersion in an aqueous 5% HF solution for 30 min at room temperature. For the preparation of membrane photoelectrodes, the Ta2O5 NT membranes were coated with a Ti nanoxide paste (Solaronix SA) on Ti foils by doctor blading, followed by annealing in a NH3 atmosphere at 950 °C to convert to Ta3N5/TiNx (TiNx from Ti nanoxide). The temperature was ramped up with a heating rate of 10 °C min-1 , kept at the desired temperature for 1 h, and finally the furnace was cooled down to room temperature.

The photoelectrochemical experiments were carried out under simulated AM 1.5G (100mW cm-2 ) illumination provided by a solar simulator (300 W Xe with optical filter, Solarlight; RT). 1 M KOH aqueous solution was used as an electrolyte. The Ta3N5 layers were prior to measurements coated with a Co(OH) x co-catalyst as described in [15]. Photocurrent vs.voltage (I-V) characteristics were recorded by scanning the potential from 1.0 V to 0.6 V (vs. Ag/AgCl (3 M KCl)) with a scan rate of 10 mV s -1 using a Jaissle IMP 88 PC potentiostat. Electrochemical impedance spectroscopy (EIS) measurements were performed using a Zahner IM6 (Zahner Elektrik, Kronach, Germany). EIS measurements were performed by applying 1.23 V RHE at a frequency range of 105 Hz to 0.01 Hz with an amplitude of 10 mV in the dark. Solid-state conductivity measurements were carried out in an adapted scanning electron microscope (SEM, TESCAN LYRA3 XMU) by 2-point measurements using a semiconductor characterization system (Keithley 4200-SCS). Tungsten tips were used as electrical contacts between tube tops and substrate. Resistivity values were then obtained from I-V curves with 5 mV/s sweep rate.

A field-emission scanning electrode microscope (Hitachi FE-SEM S4800, Japan) was used for the morphological characterization of the electrodes. X-ray diffraction (X'pert Philips MPD with a Panalytical X'celerator detector, Germany) was carried out using graphite monochromized Cu Kα radiation (Wavelength 0.154056 nm).

Chemical characterization was carried out by X-ray photoelectron spectroscopy (PHI 5600, spectrometer, USA) using AlKα monochromatized radiation, peaks were calibrated to C1s 284.8 eV.

## 3. Results and discussion

$Ta_2O_5$ nanotube (NT) membranes were grown by electrochemical anodization of Ta, and further proceeded to membranes as described in the experimental part. The resulting both-end open NT-layers consist of aligned nanotubes, with an individual diameter of ~50 nm and a length of 10-12 μm (Fig. 1a-c). The thickness of such $Ta_2O_5$ NT membranes can be adjusted between 10 to 20 μm, by increasing the anodization time from 5 to 15 min (~15 μm and ~20 μm for 10 min and 15 min, respectively). Nevertheless, membranes with a thickness lower than 7 μm cannot be separated from the Ta substrate without breakage. The oxide membranes were then connected by doctor blading to a $TiO_2$ nanoparticle layer (~10 μm) on a Ti foil ($Ta_2O_5$ /$TiO_2$ , Fig. 1d-f); in such electrodes, ~10 nm diameter $TiO_2$ nanoparticles can be observed underneath the NT membrane layer (Fig. 1d,e) and in between the nanotubes (Fig. 1f). The $Ta_2O_5$/$TiO_2$ structure is then subjected to nitridation at 950 °C (optimized conditions) and a $Ta_3N_5$/$TiN_x$ electrode is obtained (Fig. 1g-i). The nanotube diameter decreases to ~40 nm as well as the thickness, to ~8 μm, due to the volume decrease when converting $Ta_2O_5$ to $Ta_3N_5$. Additionally, well-connected nano-sized porous $TiN_x$ particles are observed underneath the NT membrane (Fig. 1h,i). The thickness of the $TiN_x$ layer is about 10 μm (inset Fig. 1i). The $TiN_x$ nanoparticles and Ti foil act as a back contact and a conductive substrate for the Ta3N5 membrane, respectively.

To determine the phase and chemical composition of the as-formed and nitrided nanotubes or membranes, XRD and XPS investigations were performed. Fig. 2a shows the XRD patterns of Ta3N5 NT on Ta foil (Ta3N5 NT) and the membrane with TiNx back contact layer (Ta3N5/TiNx ). Both patterns show monoclinic Ta3N5 phases, while for the Ta3N5/TiNx electrode peaks corresponding to TiN and Ti2N are also observed. For Ta3N5 NT, in addition, a Ta 4 N peak at 41° and 42° can be identified, that is generally attributed to over-nitridation [7]. XPS measurements of the Ta2O5/Ta3N5 nanotubes or membranes are listed in Fig. 2b-f. The Ta 4f 7/2 peak is shifted from 27 eV for Ta2O5 to 25 eV for Ta3N5 (Fig. 2b), and for both nitrided nanotubes on Ta or Ti foil, clear N 1s peaks at 397 eV are observed (Fig. 2c). In addition, for the TiN x back contact layer, the Ti 2p and N 1s peaks (Fig 2d, f) after nitridation clearly indicate the presence of the TiN phase – see the Ti 2p peak fitting in Fig. 2e. Please note that at the top surface due to the contact with air, TiO2 and very small amounts of TiON phase are also detected.

Fig. 3a,b shows the transient photocurrent-potential curves of Ta3N5 NT membrane on Ti foil (Ta3N5/TiNx ) and for comparison purposes, Ta3N5 NTs on a Ta foil. The photocurrent of the TiN x supported electrode increases significantly at 0.6 V RHE and reaches 5.3 mA cm -2 at 1.2 V RHE (Fig. 3b) which is consistent with literature reported values for 1D nanostructures. More importantly, an anodic photoresponse is observed at potentials below 0.2 V RHE (whereas in literature this occurs at ~0.5 [7,15,17,18]), which indicates a preferable band alignment in the electrode to reduce the onset for PEC oxygen evolution from water. Additionally, we investigated the influence of the nanotube membrane thickness on the photoresponse of Ta3N5 electrodes as shown in Fig. 3c. The higher thicknesses of NT membranes, i.e. ~15 μm and ~20 μm, lead to a lower photocurrent due to an increased recombination of charge carriers. However, the ~15 μm NT membrane (anodization at 60 V for 10 min) still shows a low onset potential of 0 V RHE.

The electrical properties of nanotubes and membrane were evaluated by two-point conductivity measurements and show a 5 times decrease in the resistance of the layers and an increased electric conductivity of the membrane on the TiNx back contact layer (Fig. 3d). Please note that TiNx shows virtual metallic conductivity compared to the TaNx phase (inset of Fig. 3d). Moreover, electrochemical impedance spectroscopy measurements confirm a lower charge transfer resistance of the membrane on TiNx (Rct 10170 Ω) compared to the nanotube (Rct 50230 Ω), see Fig. 3e. These findings are in line with the work function of TiNx that varies from 3.5 and 4.4 eV [22], and thus corresponds well to the flat band potential of the Ta3N5 which is at 0.2 V RHE . Therefore, it can be concluded that TiNx forms an energetically suitable ohmic contact with Ta3N5 at the interface. The above results thus show that the existence of TiNx layer as back contact provides a beneficial effect on the charge transfer resistance, thus provides more efficient photoelectron-hole separation, and thus improves the PEC performance of Ta3N5 photoelectrodes [23].

## 4. Summary

In the current work we successfully form a photoanode, consisting of a Ta3N5 nanotube membrane fixed on a Ti foil by means of a TiNx interface layer. The fabrication consists of a sequence of anodization, doctor-blading, and nitridation. On the TiNx back contact, Ta3N5 NT membrane photoanodes with Co(OH)x decoration show an anodic photocurrent of 5.3 mA cm-2 at 1.2 V RHE in 1 M KOH under simulated AM 1.5G illumination, and a significantly increased response at low anodic potentials. The shift in the onset of the anodic photoresponse to a potential of 0.2 V RHE is ascribed to the high conductivity and a favorable Fermi level of the TiNx – back contact (compared to other back contacts). It should be noted that many parameters of the subnitrides can further be optimized to explore the full potential of the current synthesis. However, the present results clearly indicate that a TiNx as underlayer significantly enhances the photoelectrochemical water splitting of tantalum nitride photoanodes.

**Acknowledgments**

The authors would like to acknowledge DFG SPP1615 and ERC for the financial support.

**Figure 1.** SEM top, cross sectional, and bottom images of (a-c) Ta2O5 NT membrane, (d-f) Ta2O5 NT membrane on Ti foil, and (g-i) Ta3N5 NT membrane on Ti foil. The electrode was anodized at 60 V for 5 min. Insets of a, d, and g show optical images of membranes. Inset of (i) shows the cross sectional image of TiNx subnitride layer.

**Figure 2.** (a) XRD patterns of Ta3N5 NTs on Ta foil and Ta3N NT membrane on Ti foil (Ta3N5 /TiN x ); (b-f) high resolution XPS spectra of (b) Ta4f for Ta2O5 and Ta3N5 NTs, Ta2O5/TiO2 and Ta3N5/TiN x , (c) N1s for Ta2O5 and Ta3N5 NTs, Ta3N5/TiN x , (d-f) Ti2p and N1s for TiO2, TiN x with (e) peak fitting of Ti2p for TiN x .

**Figure 3.** (a,b) PEC water splitting properties of (a) Co(OH) x treated Ta3N5 NTs and (b) the corresponding membranes on Ti (Ta3N5 /TiN x ) (nanotubes were obtained at 60 V, 5 min); (c) PEC water splitting properties of Co(OH) x decorated Ta3N5/TiN x , as a function of the starting length of the Ta2O5 membranes converted to Ta3N5 . The electrolyte is 1 M KOH under simulated AM 1.5G light (100 mW cm-2 ); (d) electrical resistance and (e) electrochemical impedance spectroscopy (EIS) of Ta3N5 NTs and Ta3N5/TiNx . Inset of (d) shows the corresponding results of TiNx and TaNx.

**Figure 1**

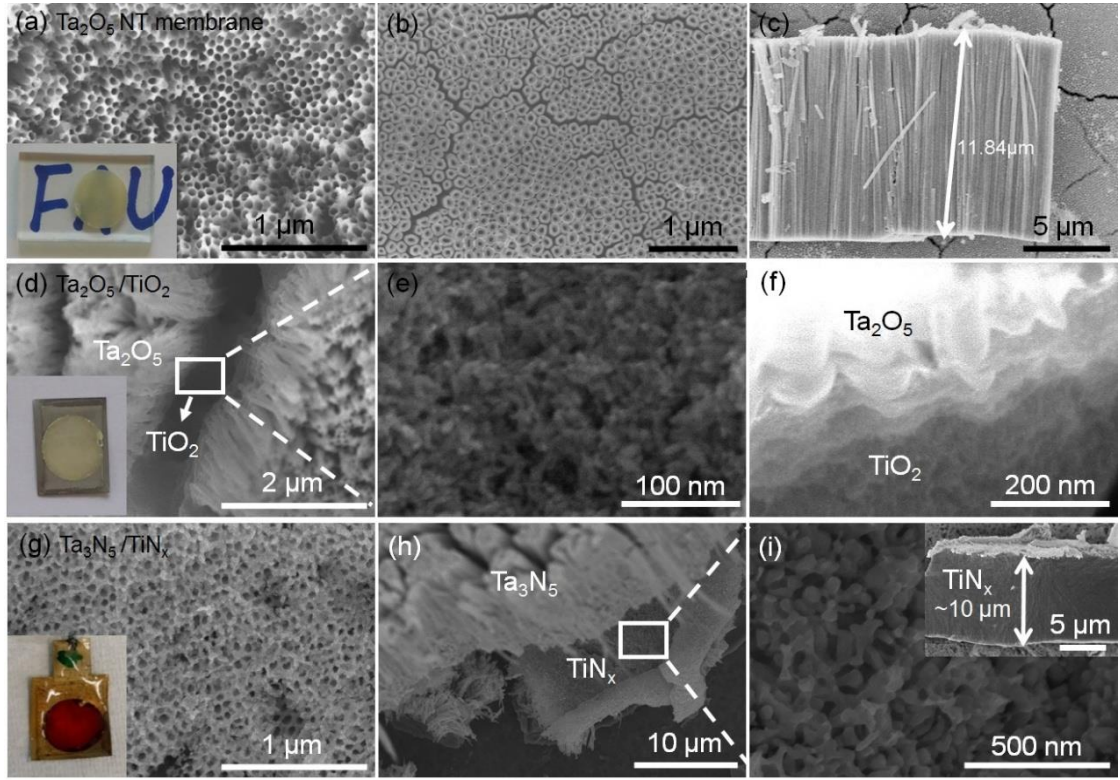

**Figure 2**

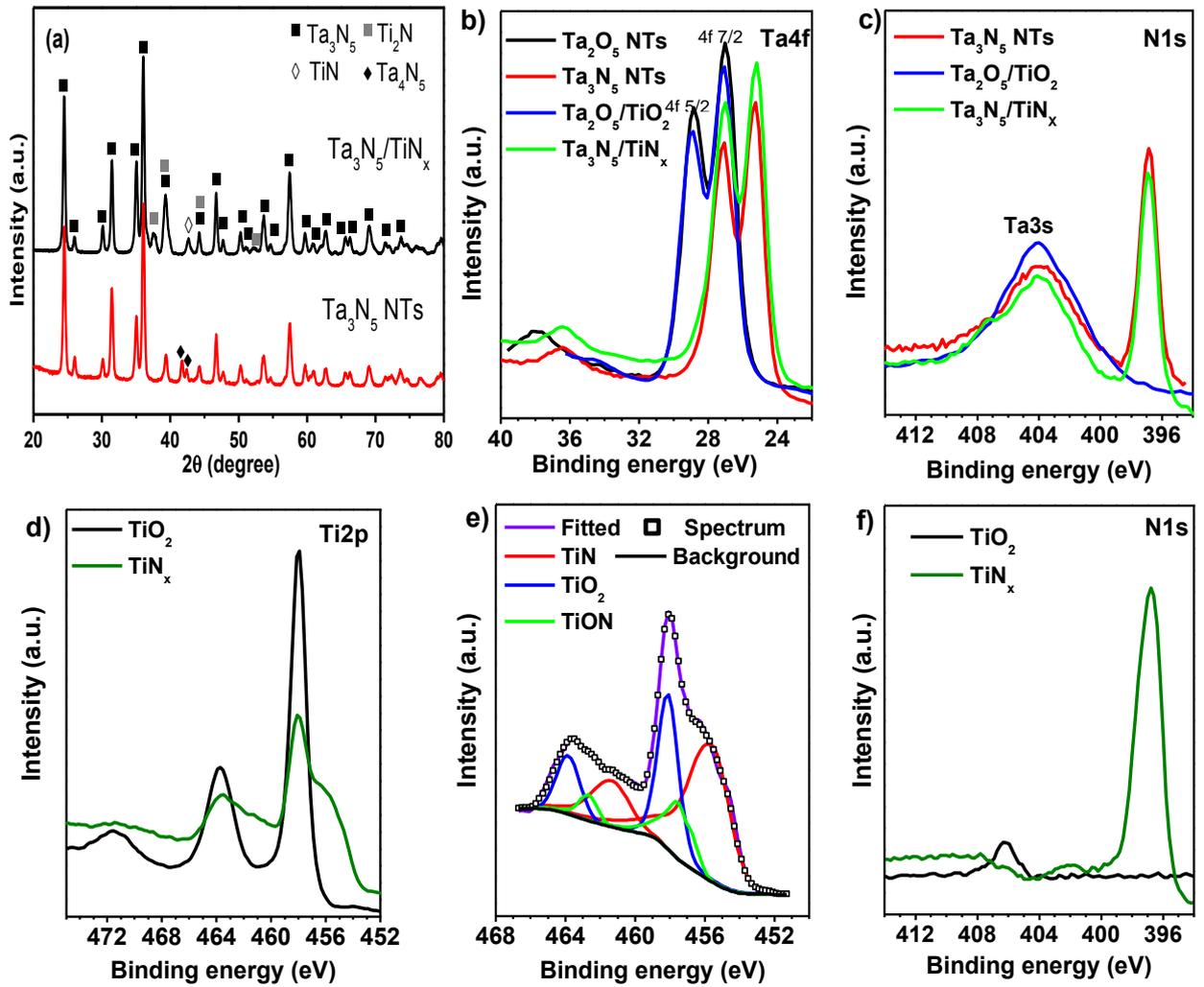

**Figure 3**

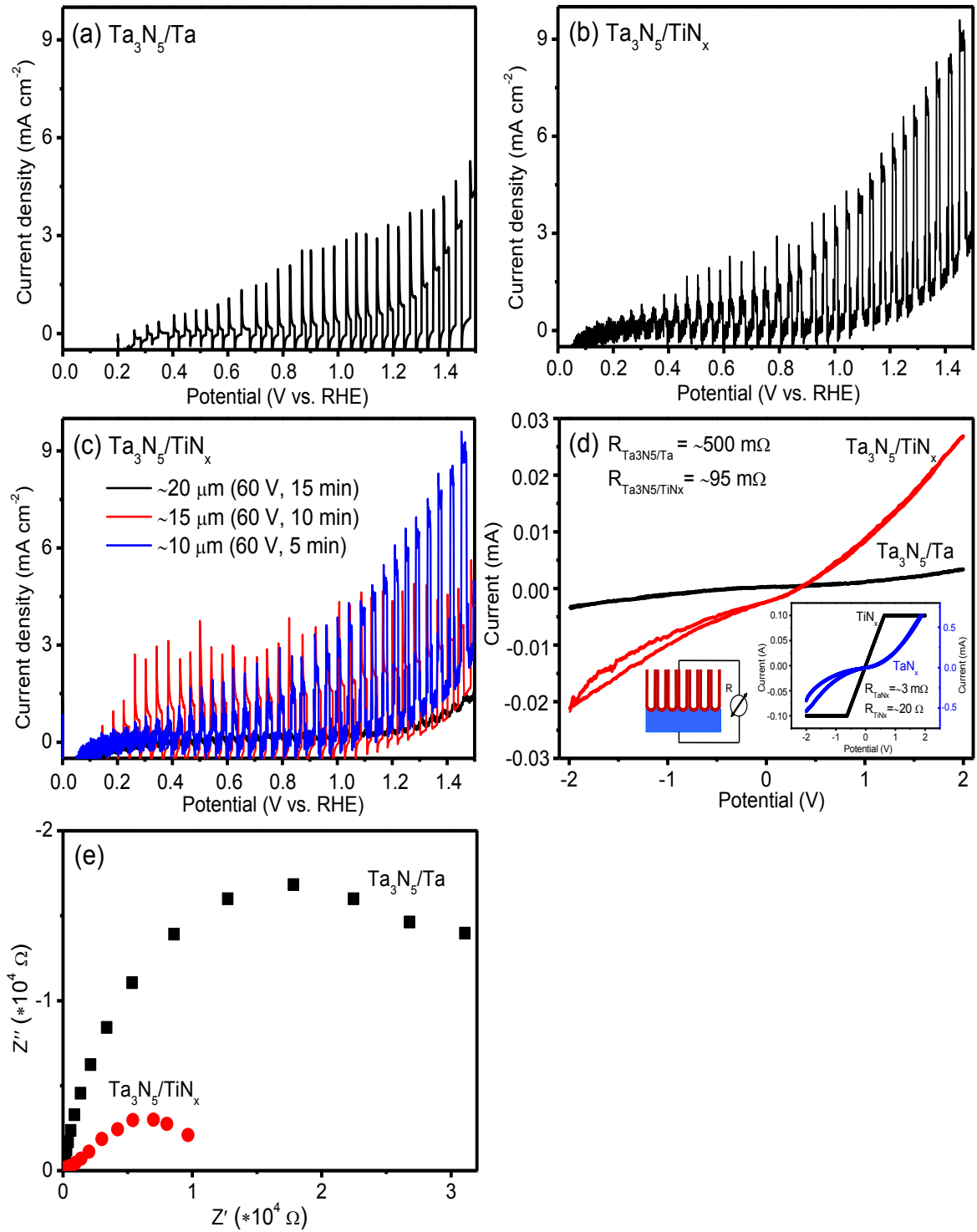